\documentclass[%
 aps,
 sd,%
 amsmath,amssymb,
 reprint,%
]{revtex4-1}

\usepackage{graphicx}
\usepackage{dcolumn}
\usepackage{bm}

\usepackage[utf8]{inputenc}

\usepackage[colorlinks]{hyperref}
\usepackage{float}

\usepackage{amsmath}

\usepackage{xspace}

\newcommand{\mvec}[1]{\ensuremath{\mathbf{#1}}}

\begin{document}

\title[]{Interfacial Dzyaloshinskii-Moriya Interaction of Antiferromagnetic Materials}

\author{Md. Rakibul Karim Akanda}
\thanks{makan001@ucr.edu}
\affiliation{Department of Electrical and Computer Engineering, University of California - Riverside, CA 92521}

\author{In Jun Park}
\affiliation{Department of Electrical and Computer Engineering, University of California - Riverside, CA 92521}

\author{Roger K. Lake}
\thanks{rlake@ece.ucr.edu}
\affiliation{Department of Electrical and Computer Engineering, University of California - Riverside, CA 92521}

\begin{abstract}

The interface between a ferromagnet (FM) or antiferromagnet (AFM) and a heavy metal (HM)
results in an antisymmetric exchange interaction known as the
interfacial Dzyaloshinskii-Moriya interaction (iDMI) which favors non-collinear spin configurations.
The iDMI is responsible for stabilizing noncollinear spin textures such as skyrmions in materials
with bulk inversion symmetry.
Interfacial DMI values have been previously determined theoretically and experimentally for FM/HM interfaces,
and, in this work, values are calculated for the metallic AFM MnPt and the insulating AFM
NiO.
The heavy metals considered are W, Re, and Au.
The effects of the AFM and HM thicknesses are determined.
The iDMI values of the MnPt heterolayers are comparable to those of the common FM materials, and those
of NiO are lower.
\end{abstract}

\maketitle

\section{Introduction}

Antiferromagnetic (AFM) materials are of current interest due to their
insensitivity to external magnetic fields,
absence of demagnetizing fields,
scalability to small dimensions,
widespread availability in nature with high N\'{e}el temperatures,
and operation in the THz frequency range.\cite{AFM1,AFM2,AFM3,AFM4,AFM5,AFM6,AFM7,AFM8,AFM9,AFM10,AFM11,AFMnew12,AFMnew13,AFMnew14,AFMnew15}
Antiferromagnetic materials exhibit various interesting phenomena such as a large anomalous Hall effect,
spin Seebeck effect, spin Hall magnetoresistance and spin galvanic effects.\cite{AFMeffect1,AFMeffect2,AFMeffect3,AFMeffect4,AFMeffect5,AFMeffect6,AFMeffect7,AFMeffect8,AFMeffect9,AFMeffect10,AFMeffect11,AFMeffect12}
They offer various promising spintronic applications exploiting skyrmions or domain walls
which can be used as non-volatile memory and racetrack memory.\cite{Application1,Application2,Application3,Application4,Application5,Application6,Application7,Application8}

Noncollinear spin configurations
such as skyrmions and chiral helices are found in non-centrosymmetric materials
with broken inversion symmetry.\cite{Noncentrosymmetric1,Noncentrosymmetric2,Noncentrosymmetric3}
Since non-centrosymmetric magnetic materials are not common in nature,
heterolayers consisting of a
heavy metal (HM), which offers high spin orbit coupling (SOC),
and a magnetic material can be created that breaks inversion symmetry at the interface.
At an HM/FM or HM/AFM interface,
the interfacial Dzyaloshinskii-Moriya interaction (iDMI),
which is an anti-symmetric exchange interaction,
can stabilize N\'{e}el type domain walls and skyrmions.\cite{iSymmetry,roomT,roomT2,roomT3,SAF2020,AFMskyrmion,DomainWallTilt,DomainWallWalker,skyrmionNew1,domainWallNew1}
The interfacial DMI of FM materials is generally measured using Brillouin light scattering
(BLS) \cite{BLS1,BLS2,BLS3,BLS4,BLS5,BLS6,AFMfm}.
Experimentally AFM spin textures are measured using spin polarized scanning tunneling microscopy (STM),
and the bulk DMI in non-centrosymmetric materials is measured using inelastic neutron scattering.\cite{STM,Neutron}

\begin{figure}
\includegraphics[width=0.4\textwidth]{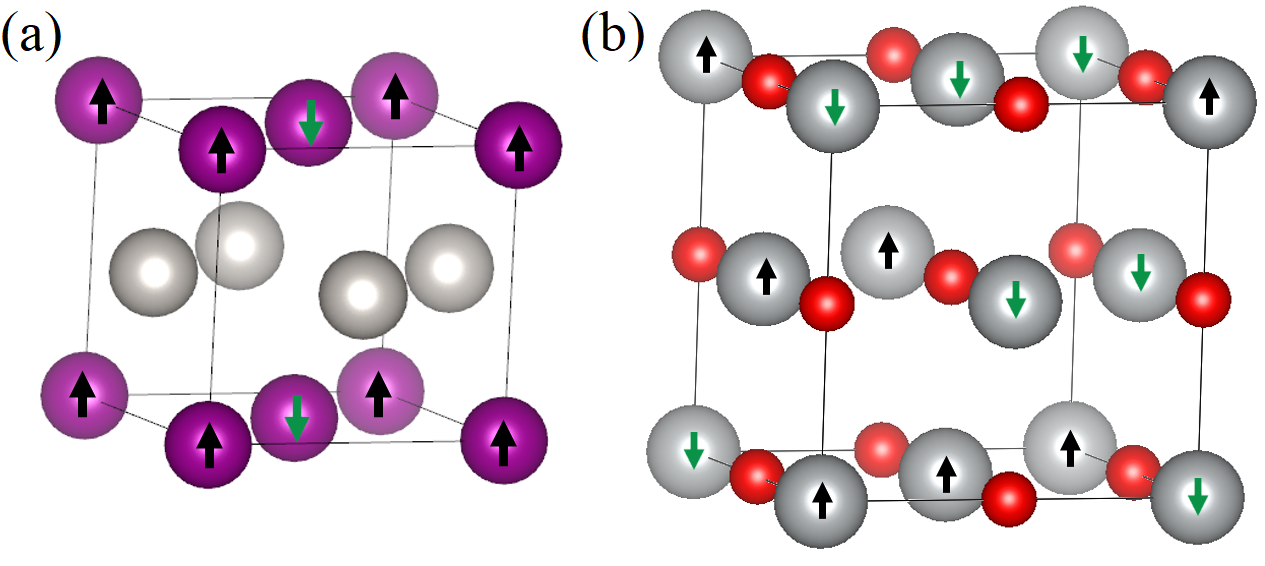}\\
\caption{(a) Unit cell of MnPt having stripe-type AFM spin configurations.
(b) Unit cell of NiO having G-type AFM spin configurations.
Spin directions are shown in 3d orbital magnetic materials (Mn or Ni) and atoms without spin are Pt (grey) and oxygen (red).}
\label{fig:MnPtNiO}
\end{figure}

To the best of our knowledge, there are no existing reports of measurements or calculations of the iDMI of
HM/AFM heterolayers.
To begin to fill that knowledge gap,
this paper describes the results of ab initio calculations of the iDMI for
three different material combinations of HMs and AFMs.
Since interfacial DMI helps to stabilize N\'eel type spin textures that rotate in the out of plane direction,
the AFM materials MnPt and NiO are chosen with N\'eel vectors oriented out of the plane.\cite{OutOfPlaneNeel1,OutOfPlane2,OutOfPlane3,OutOfPlane4,OutOfPlane5}
Both MnPt and NiO have high N\'eel temperatures (975K and 530K, respectively).\cite{NeelTemp2,NeelTemp3,NeelTemp4}
MnPt has a stripe-type AFM spin texture, and NiO has a G-type AFM spin texture.
Creating heterolayers for density functional theory (DFT) calculations
requires lattice matching to create a periodic structure.
This generally requires choosing materials with lattice constants that are relatively close to minimize strain.
For this reason, the two HMs paired with MnPt are tungsten (W) and rhenium (Re),
and the HM paired with NiO is gold (Au).
MnPt-W has lattice mismatch of 1.49\%, MnPt-Re has a mismatch of 1.25\%, and NiO-Au has a mismatch of 0.7\%.
The effects of thickness variation of the AFM layer and the heavy metal layer on the iDMI are considered.
The magnetic moments and magnetizations are also calculated as functions of layer thicknesses.
These values provide required input parameters for micromagnetic modelling of AFM spin textures.

\begin{figure}
\includegraphics[width=0.45\textwidth]{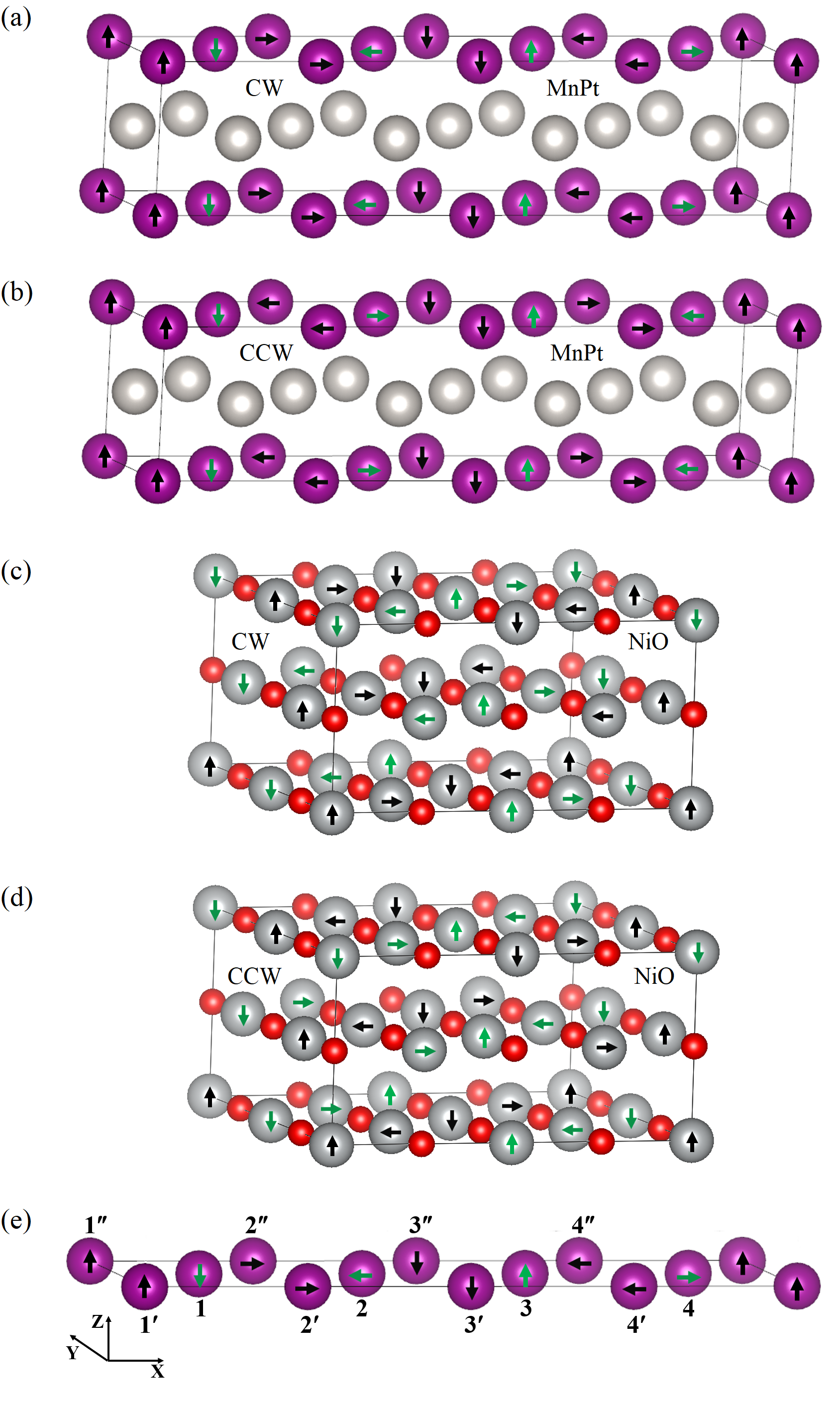}\\
\caption{Spiral spin configurations used in the DMI calculations for (a,b,e) MnPt and (c,d) NiO.
A 4 unit cell supercell of MnPt is constructed with a spin spiral in the
(a) clockwise and (b) counterclockwise directions.
A $2\times2\times1$ unit cell supercell of NiO is constructed with a spin spiral in the
(c) clockwise and
(d) counterclockwise directions.
(e) Clockwise rotated spin configuration in one layer of MnPt with labels used in Eq. (\ref{eq:Edmi_all_terms}).
}
\label{fig:spin_spirals_all}
\end{figure}

\section{Method}
The interfacial DMI is evaluated using first principles calculations based on the
Vienna Ab initio Simulation Package (VASP).\cite{VASP}
The electron-core interactions are described by the projected augmented wave (PAW) potentials,\cite{PAW}
and the exchange correlation energy is included with the generalized gradient approximation (GGA)
parameterized by Perdew-Burke-Ernzerhof (PBE).\cite{PBE}
The cutoff energy for the plane wave basis is 550 eV in all calculations.
The Monkhorst-Pack scheme is used with $\Gamma$-centered $10\times 10 \times 1$ k-point grids.

\begin{figure}
\includegraphics[width=0.5\textwidth]{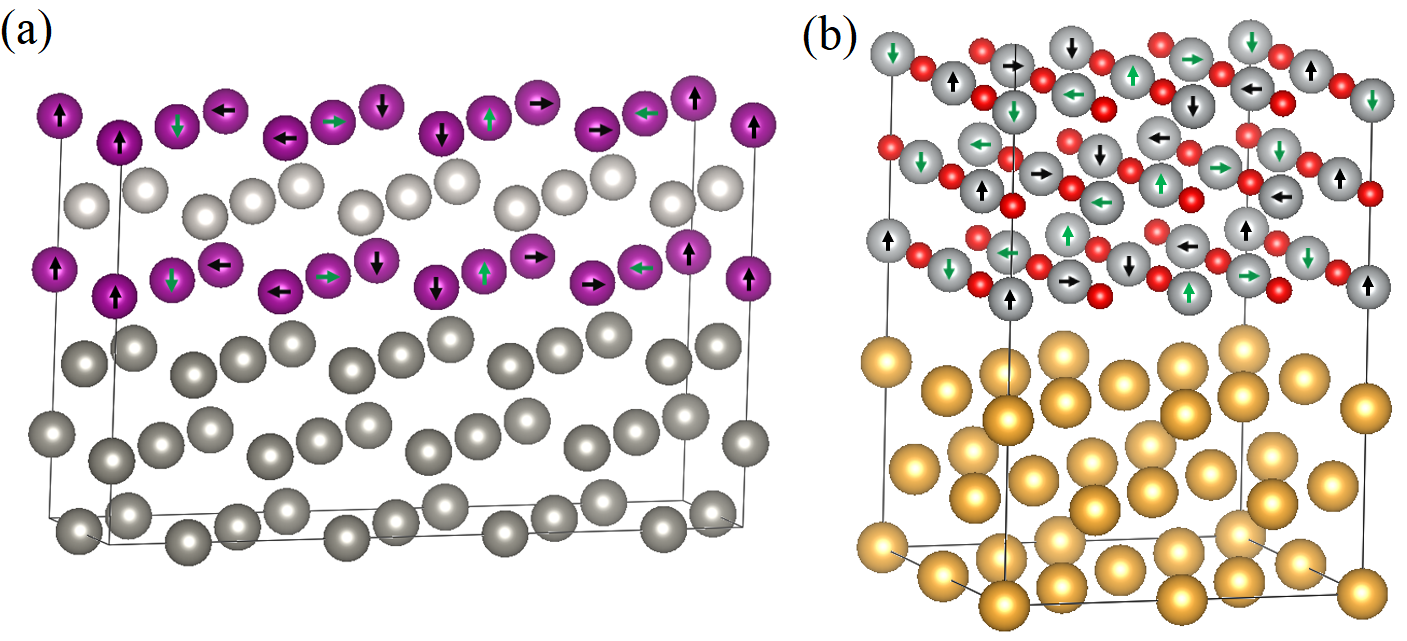}\\
\caption{(a) MnPt on top of W. Mn atoms are purple.
(b) NiO on top of Au. Oxygen atoms are red.}
\label{fig:MnPtWandNiOAu}
\end{figure}

For calculations of the iDMI,
the unit cell in Fig.~\ref{fig:MnPtNiO} is repeated in Fig.~\ref{fig:spin_spirals_all}
so that spin orientations can be gradually rotated in clockwise and anti-clockwise
directions.\cite{AnatomyCoPt,SpiralDMI2,SpiralDMI3}
MnPt and W (or Re) unit cells are repeated four times to make a supercell ($4\times1\times1$) in which
spins of the Mn atoms are rotated one period over the length of the supercell
(Fig.~\ref{fig:MnPtWandNiOAu}(a)).
NiO and Au unit cells are repeated to make a supercell ($2\times2\times1$) in which the spins of the
Ni atoms are rotated once over the length of the supercell (Fig.~\ref{fig:MnPtWandNiOAu}(b)).
Calculations are performed in three steps to find the interfacial Dzyaloshinskii-Moriya interaction.\cite{AnatomyCoPt,SpiralDMI2,SpiralDMI3}
The supercell structure is relaxed until the forces are smaller than 0.01 eV/Å
to determine the most stable interfacial geometries.
Next, the Kohn–Sham equations are solved without spin orbit coupling (SOC)
to determine the charge distribution of the system’s ground state.
Finally spin orbit coupling (SOC) is included and the self-consistent total
energy of the system is determined as a function of the orientation of the magnetic moments.
The energy difference between the counterclockwise and clockwise spin spiral is used to determine the interfacial DMI.
The total DMI strength, $D_{tot}$, is found from the energy difference between the
clockwise and anticlockwise spin configurations.\cite{AnatomyCoPt}

The energy due to the Dzyaloshinskii-Moriya interaction can be written as
\begin{equation}
E_{\rm DMI}=\sum_{\langle i,j \rangle} \mvec{d}_{ij}\cdot [\mvec{S}_i \times \mvec{S}_j]
\label{eq_ham}
\end{equation}
where $\mvec{d}_{ij}$ is the DMI vector and $\mvec{S}_i$ is the unit vector of a
magnetic moment.
$\mvec{d}_{ij}$ is found from $d(\hat{\mvec{z}}\times\hat{\mvec{u}}_{ij})$, where $\hat{\mvec{u}}_{ij}$ is the unit vector between sites $i$
and $j$, and $\hat{\mvec{z}}$ is the direction normal to the film which is oriented from heavy metal to the magnetic film.\cite{AnatomyCoPt}
Considering the four nearest neighbours in the same layer in the FCC structure, the energy of atom 2 can be written as
(see Fig.~\ref{fig:spin_spirals_all}(e)),
\begin{align}
E_2= &\frac{1}{2}[\mvec{d}_{22'}\cdot (\mvec{S}_2\times \mvec{S}_{2'})+\mvec{d}_{22''}\cdot (\mvec{S}_2\times \mvec{S}_{2''})
\nonumber \\
&+\mvec{d}_{23'}\cdot (\mvec{S}_2\times \mvec{S}_{3'})+\mvec{d}_{23''}\cdot (\mvec{S}_2\times \mvec{S}_{3''})]
\nonumber \\
&+E_{\rm other}
\label{eq:Edmi_all_terms}
\end{align}
where $E_{\rm other}$ is the spin independent, anisotropy and symmetric exchange energy contributions.
The first two terms on the right side of Eq. (\ref{eq:Edmi_all_terms})
are zero due to parallel or antiparallel magnetic moments in the cross products.
Therefore,
\begin{align}
E_2=\frac{1}{2}[\frac{1}{\sqrt{2}}D_{tot}+\frac{1}{\sqrt{2}}D_{tot}]+E_{other}
\end{align}
where the factors of $\frac{1}{\sqrt{2}}$ are due to the $45^\circ$ angle between
$\mvec{d}_{23'}$ and $\mvec{S}_2\times \mvec{S}_{3'}$.
$D_{tot}$ is the total DMI strength which is considered as the DMI strength concentrated in a
single atomic layer.\cite{AnatomyCoPt}
The energies of atom 2 considering clockwise (CW) and anticlockwise (ACW) configurations are
\begin{equation}
E_{2,CW}=\frac{1}{\sqrt{2}}D_{tot}+E_{other}
\end{equation}
\begin{equation}
E_{2,ACW}=-\frac{1}{\sqrt{2}}D_{tot}+E_{other}
\end{equation}
A single Mn layer of a supercell that contains eight magnetic atoms is shown in
Fig.~\ref{fig:spin_spirals_all}(e).
Note that the last line of atoms along the $x$ and $y$ directions are repeated and belong to
the next supercell, and are therefore not counted.
For the Mn layer of Fig.~\ref{fig:spin_spirals_all}(e),
\begin{equation}
\Delta E_{DMI}=(E_{CW}-E_{ACW})=8\sqrt{2}D_{tot} .
\end{equation}
Finally $D_{tot}$ can be written as
\begin{equation}
D_{tot}=(E_{CW}-E_{ACW})/m,
\label{eq:Dtot}
\end{equation}
where $m=8\sqrt{2}$ which depends on the cycloid wavelength.
The energy difference between clockwise and anticlockwise orientations
is calculated from DFT and divided by $m$ to obtain $D_{tot}$.

The parameter used in micromagnetic simulations of the Landau-Lifshitz-Gilbert (LLG) equation,
which we denote as the micromagnetic DMI, is also calculated.
The micromagnetic DMI is found from the total DMI strength ($D_{tot}$).
Considering four nearest neighbours, the micromagnetic DMI is found following the procedure
of Hongxin Yang et al.\cite{AnatomyCoPt},
\begin{equation}
D=\frac{4D_{tot}}{N_F a^2},
\end{equation}
where $a$ is the lattice constant, and $N_F$ is the number of magnetic layers.
The magnetic moment and magnetization are also calculated from the collinear AFM spin configurations
of the three different combinations of materials where the magnetization is the magnetic moment per unit volume.

\begin{figure}
\includegraphics[width=0.5\textwidth]{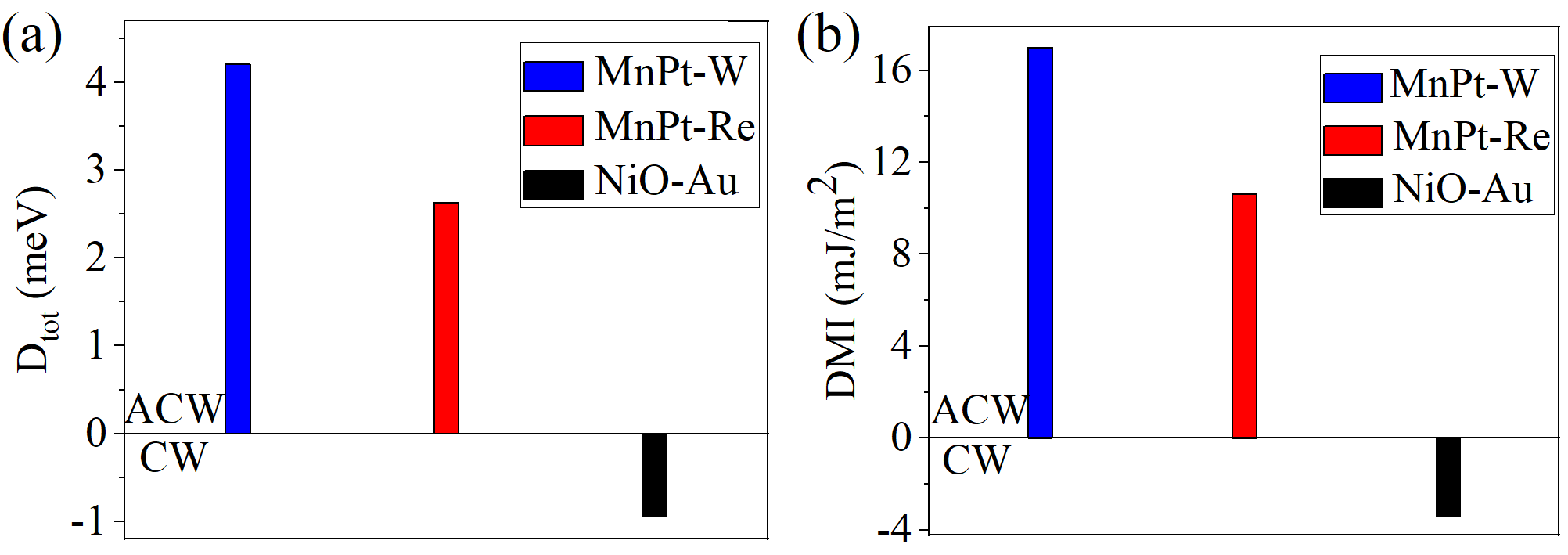}\\
\caption{
Interfacial Dzyaloshinskii-Moriya interaction of MnPt-W, MnPt-Re, NiO-Au using three layers of HM and one AFM layer:
(a) D\textsubscript{tot} per 3d orbital magnetic atom and
(b) micromagnetic DMI.
}
\label{fig:iDMI1layer}
\end{figure}

\begin{figure}
\includegraphics[width=0.5\textwidth]{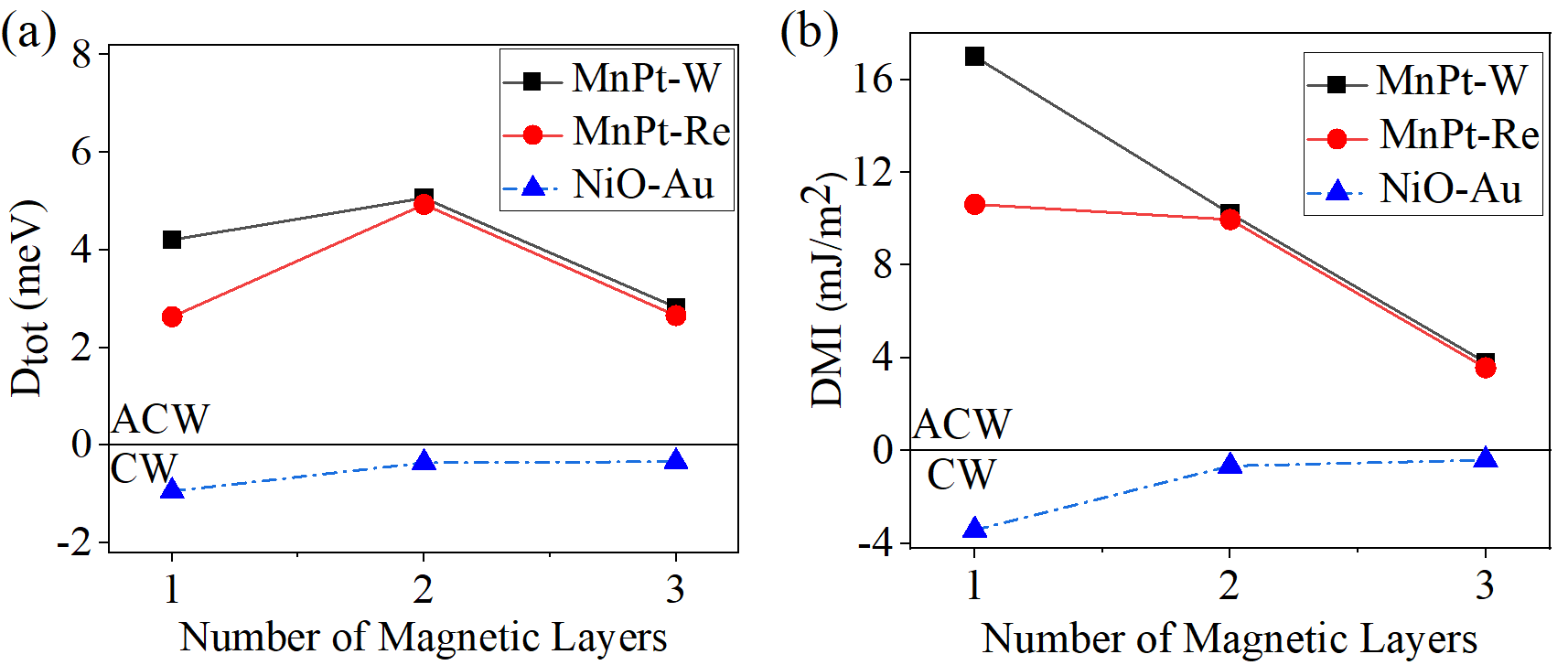}\\
	\caption{Change of (a) D\textsubscript{tot} and (b) micromagnetic DMI of MnPt-W, MnPt-Re and NiO-Au with the thickness variation of magnetic layer.
The thickness of the HM is fixed at 3 layers.
}\label{fig:DtotDMIMnPtWReNiO123Layers}
\end{figure}

\begin{figure}
\includegraphics[width=0.5\textwidth]{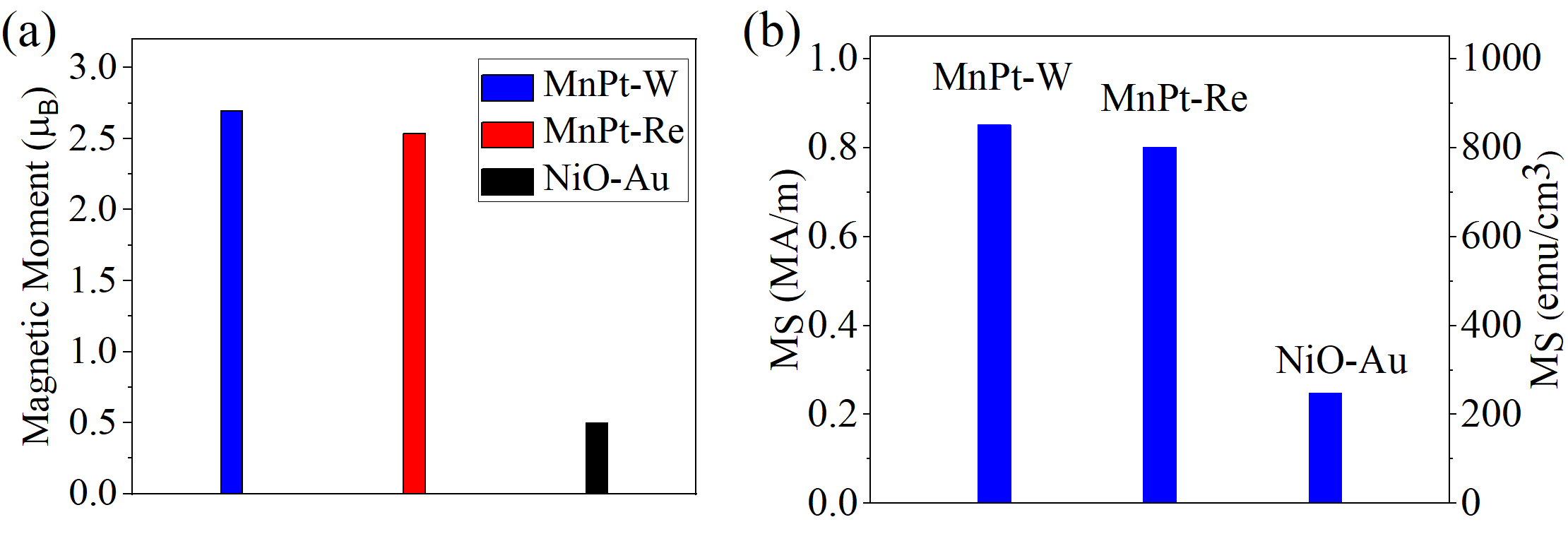}\\
\caption{(a) Magnetic moments of 3d transition metal atoms in MnPt-W, MnPt-Re and NiO-Au
using a single layer of AFM on top of three layers of heavy metal.
(b) Satruation magnetization of AFM materials.}
\label{fig:Bohr1Ms1}
\end{figure}

\begin{figure}
\includegraphics[width=0.5\textwidth]{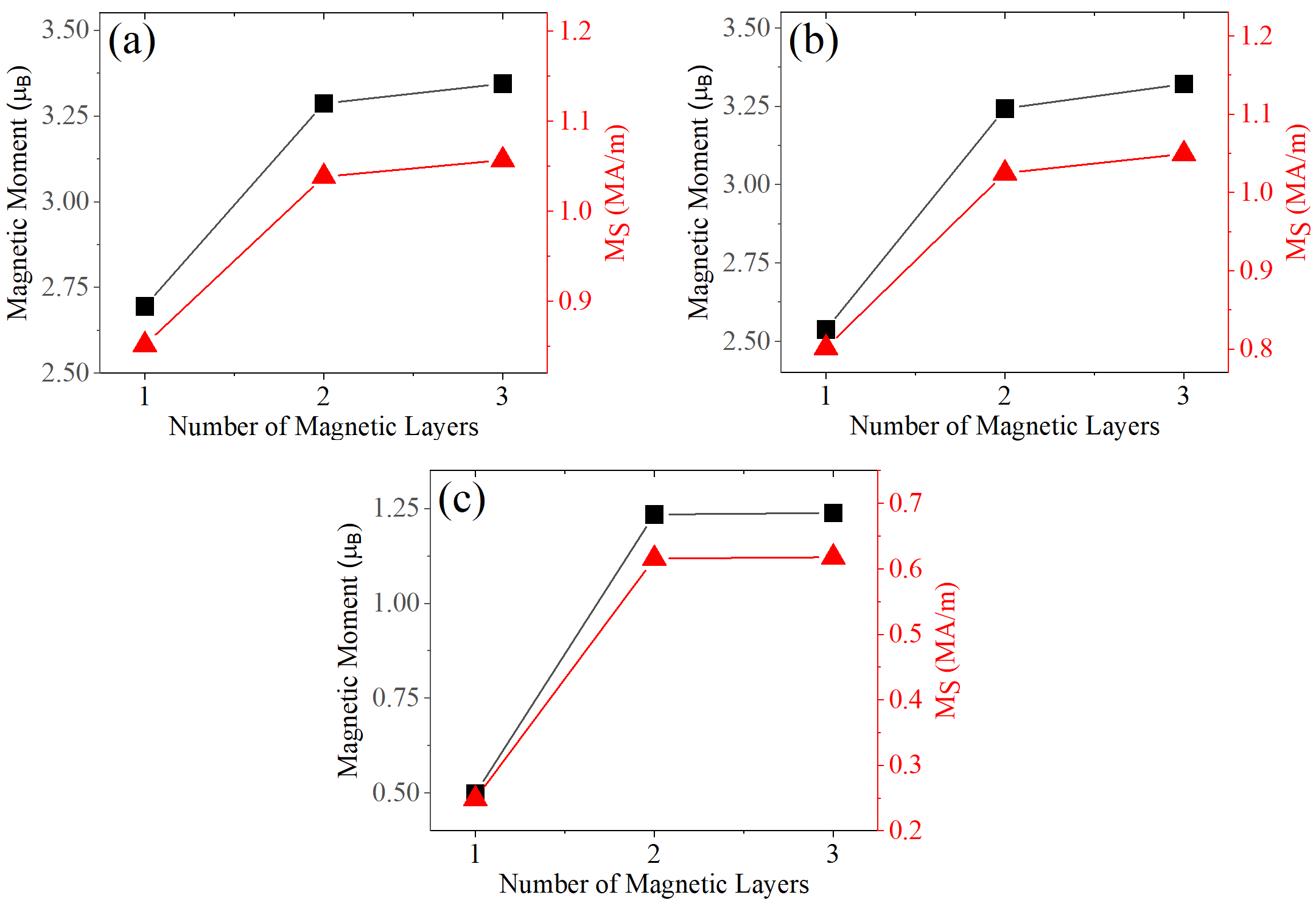}\\
\caption{Change of magnetic moments and magnetization of (a) MnPt-W, (b) MnPt-Re and (c) NiO-Au with thickness variation of the magnetic material.
The thickness of the HM is fixed at 3 layers.
Increasing the thickness increases the magnetic moments towards their bulk values.}
\label{fig:BohrMnPtWReNiOAu3layers}
\end{figure}

\begin{figure}
\includegraphics[width=0.5\textwidth]{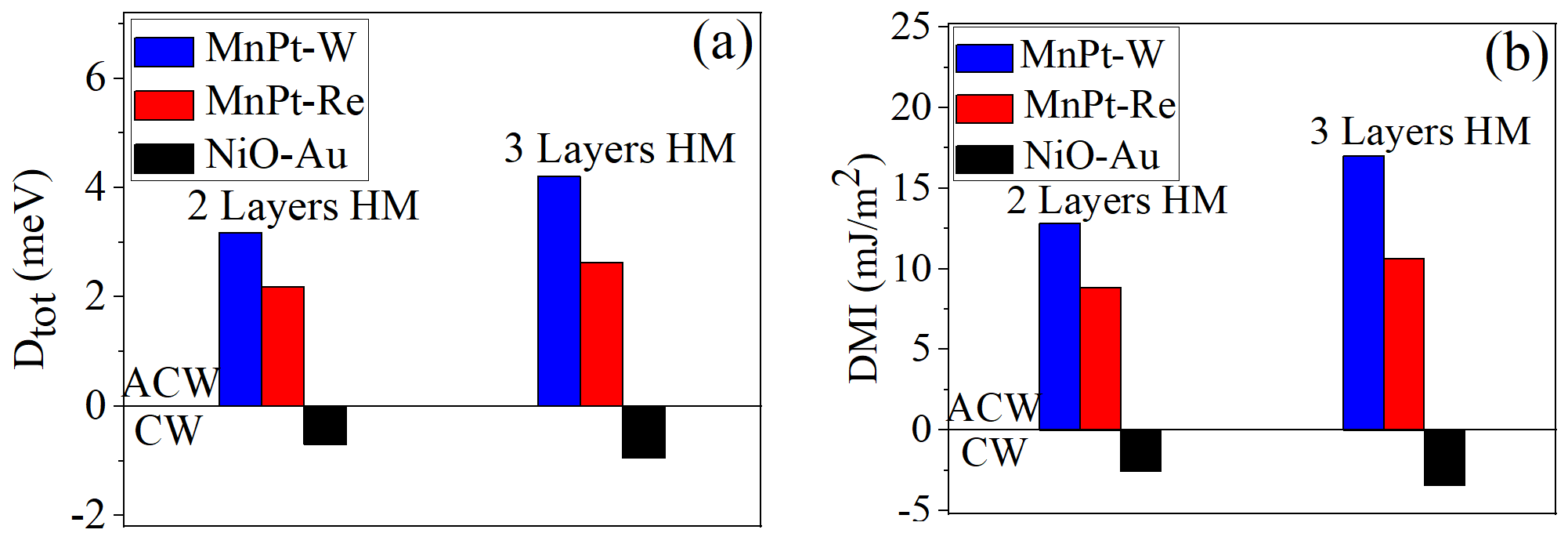}\\
\caption{
Change of interfacial DMI of monolayer AFM with thickness variation of heavy metal layer: (a) D\textsubscript{tot} with one magnetic layer and
(b) micromagnetic DMI.
Increasing the thickness of heavy metal provides more spin orbit coupling and increases the interfacial DMI.
}
\label{fig:metalThicknessDtotDMI}
\end{figure}

\section{Results}
\label{sec:iDMI-Results}
The iDMI of MnPt-Re, MnPt-W, and NiO-Au are shown in Fig. \ref{fig:iDMI1layer}.
The structures consist of one layer of AFM and three layers of HM.
The trends and quantitative values are similar to those in HM/FM interfaces \cite{Hund}.
%
%
For a monolayer of MnPt, the strength of the iDMI, in
terms of $D_{tot}$, is larger with W than with Re,
and this is consistent with the fact that the SOC of W is larger than that of Re.
Both MnPt-W and MnPt-Re show larger DMI compared to that of NiO-Au,
and this is consistent with the fact that the magnetic moment of Mn
is larger than that of Ni.
The trend that
higher magnetic moments provide larger interfacial DMI following Hund’s rule \cite{Hund}
is also observed in HM/FM interfaces.

Once the thickness of the AFM material is more than a few layers,
the volume averaged interfacial DMI decreases with increasing thickness of the magnetic material,
and it will eventually approach zero for thick magnetic material.\cite{AnatomyCoPt,FMThickness,FMThickness}
However, this does not necessarily hold true for the first two to three layers
in terms of the total DMI strength ($D_{tot}$).
For example, in MnPt, each layer of magnetic atoms (Mn) is sandwiched between layers of HM atoms (Pt).
Thus, the total DMI strength ($D_{tot}$) depends not only on the spin orbit coupling (SOC)
arising from the interfacial heavy metal W (or Re),
but also on the SOC from the inner heavy metal Pt,
and the resulting total DMI strength initially varies non-monotonically
with layer number as shown in Fig. \ref{fig:DtotDMIMnPtWReNiO123Layers}.
In contrast, NiO contains no HM atoms, and the magnitude of total DMI strength ($D_{tot}$)
decreases smoothly with thickness as shown in Fig. \ref{fig:DtotDMIMnPtWReNiO123Layers}.

The interfacial DMI values of antiferromagnetic materials are comparable to those of ferromagnetic materials.
The iDMI of 3 layers of MnPt on W of 3.8 mJ/m$^2$ is comparable to that of Co on Pt \cite{AnatomyCoPt,SpiralDMI2}.
Three layers of Ni with graphene have an iDMI of approximately 0.5 mJ/m\textsuperscript{2}\cite{SpiralDMI3},
and three layers of AFM NiO on Au has an iDMI of 0.4 mJ/m\textsuperscript{2}.
The reduction of iDMI in NiO is due to the presence of oxygen atoms which cannot directly contribute towards
the iDMI.

Since the magnitude of the iDMI depends on local magnetic moments,
comparisons of the magnetic moments of monolayer MnPt and NiO with three layers of heavy metal are
shown in Fig. \ref{fig:Bohr1Ms1}(a),
and the saturation magnetization is shown in Fig.~\ref{fig:Bohr1Ms1}(b).
NiO-Au has smaller magnetic moments which is one of the factor for its lower interfacial DMI compared to MnPt-W and MnPt-Re.
The magnetic moments are determined from the collinear AFM spin configuration.
The values are smaller than those of the bulk.
There are two reasons for smaller magnetic moments:
(i) the magnetic moments decrease with the reduction of
thickness of the magnetic materials\cite{SrRuO3,MomentThickness}
and
(ii) bonding at the AFM-HM interface due to orbital hybridization.\cite{Hund}
The HM layer changes the magnetic moments of AFM due to the hybridization
between 3d and 5d orbitals and thus changes iDMI.\cite{HM1,HM2}
The magnetic moments increase with increasing thicknesses of the AFM thin-films,
as shown in  Fig.~\ref{fig:BohrMnPtWReNiOAu3layers}, and they
approach the bulk values of MnPt and NiO.\cite{MnPtMoment1,MnPtMoment2,NiOMoment}
Thick AFM material will have larger magnetic moments
but smaller average SOC from the adjacent heavy metal.

A thicker heavy metal layer increases the iDMI, since increasing the thickness of the
heavy metal layer initially increases proximity spin orbit coupling in the AFM.
The increase in the iDMI with thickness of the HM layer is shown in Fig.  \ref{fig:metalThicknessDtotDMI}.
Since the HM provides proximity spin orbit coupling, increasing the HM thickness from one layer to several layers
increases the iDMI before it saturates after 3 or 4 layers. \cite{DMISaturate}

Besides the magnetic moments of the AFM layer, there is also a proximity induced magnetic moment in the
HM which plays an important role in determining D\textsubscript{tot}.
In FM/HM systems, a higher proximity induced magnetic moment in the heavy metal reduces the
iDMI,\cite{AnatomyCoPt} and this is consistent with what we observe with the AFM/HM system.
One layer of MnPt on three layers of Re induces a proximity magnetic moment of 0.211 Bohr magneton
on the first layer of Re atoms,
and one layer of MnPt on three layers of W induces a proximity magnetic moment of 0.206 Bohr magnetons
on the first layer of W atoms.
At the heavy metal interface, the proximity induced magnetic moment results from the
bonding of the 3d orbitals of the Mn or Ni with the 5d orbitals of the HM,
which enhances the magnetic moment of the HM atoms.

Intrinsic MnPt contains heavy metal layers of Pt sandwiched between layers of magnetic Mn atoms.
Due to the high spin orbit coupling of the inner Pt HM layer,
the antiferromagnetic MnPt slabs show a small amount of DMI even without a proximity HM layer.
For three layers of MnPt, D\textsubscript{tot} is -1.02 meV/3d atom,
and the micromagnetic DMI is -1.37 mJ/m\textsuperscript{2}.
In contrast, NiO does not have any heavy metal and does not show any DMI value without an adjacent heavy metal layer.

\begin{figure}
\includegraphics[width=0.5\textwidth]{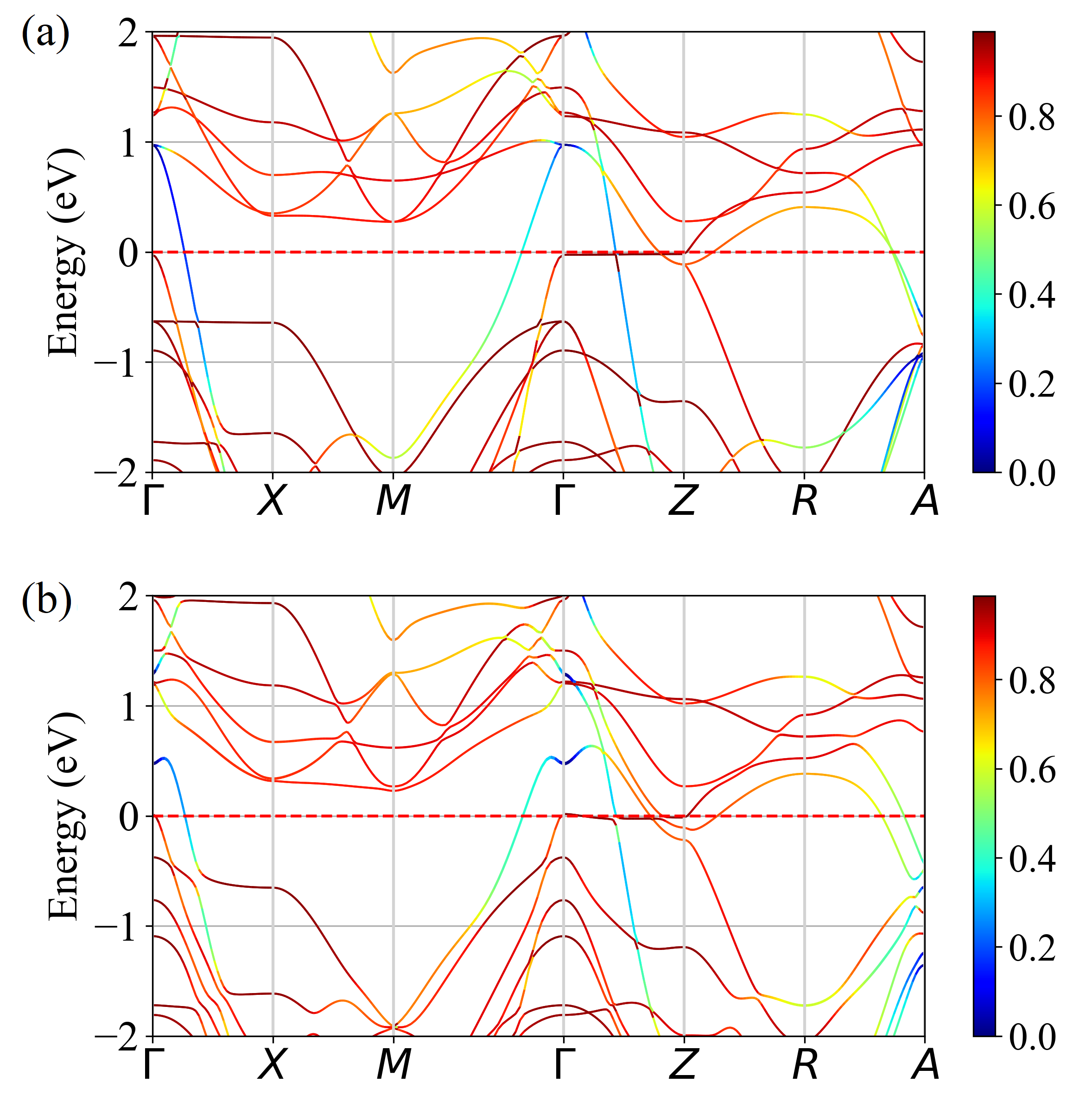}
\caption{
Band structure of MnPt:
(a) contribution of d-orbital without SOC,
(b) contribution of d-orbital with SOC.
The color bar indicates the normalized contribution from the d-orbitals
(e.g. 0.8 in the color bar represents 80\% contribution from d-orbitals).
}
\label{fig:BandStructure}
\end{figure}

Plots of the electronic band structure without and with SOC shown in Fig. \ref{fig:BandStructure}
illustrate the strength of the SOC in MnPt.
Fig. \ref{fig:BandStructure}(a) shows the bandstructure without SOC and Fig. \ref{fig:BandStructure}(b)
shows the bandstructure with SOC.
The color gives the d-orbital component of the wavefunction for each band and wavevector
as indicated by the color scale bar.
The inclusion of SOC breaks the degeneracy of the bands at $\Gamma$ resulting in band splitting of up
to 0.73 eV.

\section{Conclusions}

The interfacial Dzyaloshinskii-Moriya interaction of
MnPt/W, MnPt/Re, and NiO/Au are
calculated for different thicknesses of both the AFM and the HM.
Values of iDMI and magnetic moments required for the micromagnetic simulations
of these material combinations are determined.
The values of the iDMI of the MnPt/HM heterolayers are comparable to those of
the common ferromagnetic materials such as Fe, Co or CoFeB.
The iDMI of the NiO/Au system is approximately a factor of 7 lower.
In general, iDMI is maximized by
choosing AFM materials with larger magnetic moments,
heavy metals with high spin orbit coupling, thinner AFM layers from 1 - 3 monolayers, and HM thicknesses of
at least 3 to 4 layers.
Few layer AFM alloys that contain both HM atoms and magnetic atoms such as MnPt show an intrinsic DMI
albeit much smaller than the iDMI induced by a proximity HM layer.

These results and the quantitative values provided
will help guide experimental realization and provide needed parameters for micromagnetic
simulations of AFM materials with domain walls and skyrmions supported by iDMI.

\section{Acknowledgement}

This work was supported as part of Spins and Heat in Nanoscale Electronic Systems (SHINES) an Energy Frontier Research Center funded by the U.S. Department of Energy, Office of Science, Basic Energy Sciences under Award \#DE-SC0012670.
This work used the Extreme Science and Engineering Discovery Environment (XSEDE)\cite{towns2014xsede}, which is supported by National Science Foundation Grant No. ACI-1548562 and allocation ID TG-DMR130081.

\nocite{*}
\bibliography{references}

\end{document}